# Design of a Hard X-ray Polarimeter: *X-Calibur*

Qingzhen Guo, Alfred Garson, Matthias Beilicke, Jerrad Martin, Kuen Lee, Henric Krawczynski

*Abstract*–We report on Monte Carlo studies of the hard X-ray polarimeter *X-Calibur*. The polarimeter will be used in the focal plane of a grazing incidence hard X-ray telescope. It combines a low-Z Compton scatterer with a high-Z Cadmium Zinc Telluride (CZT) detector assembly to measure the polarization of 10 keV - 80 keV X-rays. *X-Calibur* makes use of the fact that polarized photons Compton scatter preferentially perpendicular to the electric field orientation. In contrast of competing designs, which use only a small fraction of the incoming X-rays, *X-Calibur* achieves a high detection efficiency of order unity. In this contributions, we discuss a Monte Carlo study which compares *X-Calibur's* polarimeteric performance achieved using different scattering materials (Scintillator, Be, LiH, Li), and calculate the sensitivity of *X-Calibur* when used with different balloon-borne and space-borne mirror assemblies.

## I. Scientific Motivation

X-rays are uniquely suited to study compact objects like neutron stars, pulsars, soft gamma-ray repeaters, and the emission from binary black hole systems, as these objects are very bright in the X-ray energy band. Furthermore, X-rays are well suited to study the highly relativistic outflows from gamma-ray bursts, accreting compact objects, and Active Galactic Nuclei. Polarization measurements which are based on the photoelectric effect, the Compton effect, or pair production are one of the most exciting frontiers in contemporary astrophysics, since they offer new diagnostics to discriminate between different models invoked to explain the X-ray emission from these sources. X-ray spectro-polarimetric observations increase the parameter space to study these objects from two (time variability and energy spectra) to four, by adding two qualitatively new parameters: the degree and direction of the linear polarization.

Hard X-ray observations make it possible to make substantial contributions to solving a number of important astrophysical questions, see [1] for a detailed discussion. The science drivers of hard X-ray astronomy include the following highlights:

• Combining soft and hard X-ray polarimetry, the inclination angle of binary black hole systems and the masses and spins of the central black holes can be measured or constrained [2,3]. Contemporaneous observations over the broadest possible energy range are required to derive the best possible constraints while testing at the same time the validity of the models used for fitting the observed distributions.

• Combined soft and hard X-ray observations can be used to study black hole accretion disks, and models of accretion disk coronae [3].

• Hard X-ray polarization measurements give unique information about the particle acceleration processes in pulsars, pulsar wind nebulae, and magnetars. As high-energy electrons loose their energy more rapidly than low-energy electrons, hard X-rays are likely to be produced in more compact regions than soft X-rays. As a consequence, higher polarization degrees are expected with a higher diagnostic potential.

• Spectro-polarimetric observations of X-rays from the relativistic jets from Gamma-Ray Bursts (GRBs) and Active Galactic Nuclei have the potential to establish a helical structure of the jet magnetic field in the very inner core of the jets. If the jet magnetic field is indeed helical, this would be an important confirmation of magnetic models of jet formation, acceleration, and confinement [4].

• Test models of Lorentz Invariance violations which predict a helicity dependence of the speed of light with unprecedented accuracy [5].

For general reviews on the science drivers and techniques of X-ray polarimetry, the interested reader can consult [1,6,7,8].

In 2014, NASA will launch the Gravity and Extreme Magnetism Small Explorer (GEMS), a satellite-borne X-ray polarimeter targeting the soft X-ray regime (primary energy range: 2 keV - 10 keV). For a deep observation ($10^6$ s) GEMS can measure very low polarization levels (~3%) even for weak sources (~ 2 milliCrab). The Soft Gamma-ray Imager on ASTRO-H [9] to be launched in 2013 will also have polarization sensitivity.

A GEMS follow up-mission could improve in one or more aspects on GEMS, e.g.:
• Improved sensitivity in the GEMS core energy regime from 2 keV to 10 keV;
• A wider energy bandpass, extending the spectro-polarimetric observations to lower and to higher energies;
• Improved energy resolution;
• Spectro-polarimetric imaging capabilities.
• A large field of view polarimeter would make it possible to measure the polarization of transient sources, such as GRBs [10].

In this paper, we present results from Monte Carlo studies of the polarimetric performance of *X-Calibur* (Fig. 1), a hard X-ray polarimeter sensitive from 10 keV to 80 keV [1]. *X-Calibur* is a Compton polarimeter using low-Z and high-Z materials to scatter and absorb incident X-rays. In this study, we investigate the effects of different scattering materials on *X-Calibur's* polarimetric sensitivity. For a strong Crab-like

---

Manuscript received November 21, 2010. This work was supported in part by the NASA under grant NNX10AJ56G and discretionary founding from the McDonnell Center for the Space Sciences to build the *X-Calibur* polarimeter and in part by the Chinese Scholarship Council (NO.2009629064).

Q. Guo, A. Garson, M. Beilicke, J. Martin, K. Lee and H. Krawczynski are with Department of Physics and the McDonnell Center for the Space Sciences, Washington University in St. Louis, 1 Brookings Drive, CB1105, St. Louis, MO 63110 USA (telephone: 314-935-8553, e-mail: krawcz@wuphys.wustl.edu).

Q. Guo is with the State Key Laboratory of Solidification Processing, Northwestern Polytechnical University, Xi'an, 710072, China.

source, a satellite-borne *X-Calibur* polarimeter could measure polarization degrees of less than 1% in a 100 ksec observation.

The rest of this paper is organized as follows. In Section II, we review the general considerations of the polarimeter. The *X-Calibur* design and polarimetric approach is introduced in Section III. The following section describes the simulations and analysis used in our study. Section V presents the results of the comparison and is followed by a summary.

## II. GENERAL CONSIDERATIONS

Many astrophysical sources such as pulsars are expected to emit polarized X-rays. In general terms, X-ray polarimeters take advantage of the dependence of the interaction cross sections on the polarization of the cosmic X-rays. If one accumulates many events from a linear polarized source and determines the azimuthal scattering/emission angle for each one, the compiled distribution will reveal a sinusoidal modulation with a 180˚ periodicity.

To use a polarimeter, the modulation of the azimuthal scattering angle distribution of a 100% linearly polarized signals needs to be determined. The modulation is known as the modulation factor, μ, defined as:

$$\mu = \frac{C_{max} - C_{min}}{C_{max} + C_{min}}, \quad (1)$$

where $C_{max}$, $C_{min}$ refer to the maximum and minimum numbers of counts in the azimuthal scattering angle histogram.

The polarimetric performance of a polarimeter in a specific application can be quantified by the Minimum Detectable Polarization (MDP) [11,1]. The MDP is the degree of linear polarization from a source that can be measured on a confidence level of 99% in a given observation time T:

$$MDP = \frac{4.29}{\mu R_{src}} \sqrt{\frac{R_{src} + R_{bg}}{T}}. \quad (2)$$

Here, $R_{src}$ and $R_{bg}$ are the source and background counting rates, and T is the integration time.

The modeling of the background, shielding, and background suppression is outside the scope of this paper. We limit our analysis to situations where the signal dominates strongly over the background ($R_{src} \gg R_{bg}$), and we assume $R_{bg} = 0$ in the following.

## III. THE *X-CALIBUR* DESIGN

The *X-Calibur* instrument is a Compton polarimeter. The detector exploits the fact that linearly polarized X-rays preferentially Compton scatter in a direction perpendicular to the orientation of the electric field vector. The azimuthal scattering angle distribution of a Compton polarimeter has peaks at angles +/- 90 degrees from the polarization direction of the incident photons.

In a measurement scenario, one can imagine event types where an incident X-ray Compton scatters once, and subsequently deposits the remainder of its energy in a different detector element so that the azimuthal scattering angle can be determined. For this approach, it is advantageous to use an instrument composed of two materials with different atomic numbers. In order to have a large cross section for Compton scattering, a low-Z material is desirable as a "scatterer". For detecting the scattered photon, a high-Z "absorber" increases the probability that the X-ray is fully absorbed. The low-Z/high-Z combination leads to a high fraction of unambiguously detected Compton events and suppresses the detection of elastically scattered neutrons which can mimic Compton events.

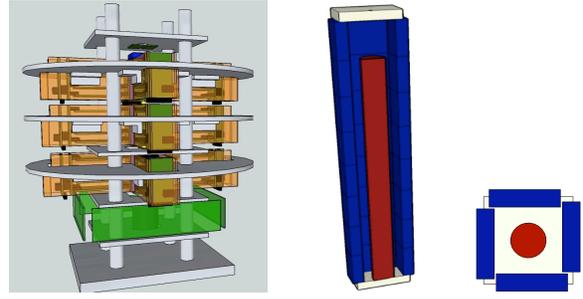

Fig. 1 Conceptual drawings of the *X-Calibur* detector assembly including the ASIC readout electronics of the CZT detectors. The PMT readout of the scintillator is not shown. The left panel shows the readout electronics surrounding the detector assembly. The middle and right panel shows the detector assembly. A Wolter mirror focuses the X-rays on the polarimeter aligned with the optical axis. X-rays Compton scatter in the low-Z scatterer (red) and are photoabsorbed in the high-Z CZT (blue). The azimuthal distribution of scattering angles constrains the linear polarization of the incoming radiation.

*X-Calibur* uses high-Z Cadmium Zinc Telluride (CZT) as the photoeffect absorber. It is the goal of this paper to compare the performance of *X-Calibur* with four different Compton scatterers (plastic scintillator, Be, LiH and LiH). The *X-Calibur* design is based on a ~0.5 cm diameter cylindrical scattering rod inside a rectangular assembly of CZT detectors (see Fig. 1). The length of the scattering rods was chosen as to yield a Compton scattering probability of ~90% for 80 keV photons. The degree of linear polarization can be measured based on events with (or without) a trigger in the scattering rod (in case that a scintillator is used) and a hit in the surrounding CZT detector. The events with a coincident trigger in the (scintillator) scattering rod and in the CZT detectors will have a much lower background than the events which trigger only one or more CZT detectors.

The CZT detector configuration is made of ~32 detector units (each 0.5×2×2 cm$^3$, more units for longer scattering rods) with a monolithic cathode oriented towards the inside of the assembly and 8×8 anode pixels (2.5 mm pitch) oriented towards the outside. The scattering rod is 1cm in diameter. The length of the CZT assembly is 2 to 3 cm longer than the scattering rod length. Table 1 lists the physical characteristics and dimensions of the different scattering rods and CZT detector assemblies for each case. We assume that the CZT detectors achieve energy thresholds 10 keV. The interaction depths of the energy depositions in the CZT detectors can be estimated based on the anode-to-cathode signal ratio [12, 13, 14]. Depth information can be used to suppress photons and charged particles that deposit their energy close to the outer edges of the CZT detector assembly.

*X-Calibur* is a focal plane instrument, to be integrated with a Wolter-type (imaging) mirror assembly similar to the one used in the *HERO* [15], *HEFT* [16], *InFOCμS* [17] and *NuStar* [16,18] experiments. Depending on the focal length and diameter of the X-ray mirrors, larger diameter rods may be required. The major axes of the scattering rod and the CZT detector assembly are aligned with the optical axis of the grazing incidence mirror assembly. When a plastic scintillator is used as the scattering material, the rod is read out with a photodetector (PMT, hybrid photodetetcor, or a Geiger mode avalanche photodiode) at the rear side of the assembly. The other three scattering materials are assumed to be passive.

The azimuthal scattering angle is determined from the position of the CZT pixel with the highest signal and by assuming that the photons scatter at the optical axis. As grazing angle mirror technology is limited to energies ≤ 80 keV, we show all results over the limited energy range from 10 keV to 80 keV. As will be shown below, the polarimeter is very sensitive. However, it does not provide imaging information even though it is located in the focal plane of a Wolter type mirror assembly.

## IV. SIMULATIONS

We do not evaluate full telescope designs in this paper but limit the discussion to the response of the detector assembly to incoming photons. We consider the particular case that the signal strongly dominates over the background and that the latter is therefore negligible. The energy range of *X-Calibur* is limited at the low-energy end by the low-energy threshold of the CZT detectors (10 keV), and at the high-energy end by the high-energy cut-off of the mirror (80 keV).

We studied the performance of the polarimeter assemblies with four different scattering materials based on a Monte Carlo study with the GEANT4 package [19]. We did the simulations with the Livermore Low-Energy Electromagnetic Models [20]. We consider balloon-borne experiment (short flight: 1 day, 6 hrs integration time; long flight: 10 days, 60 hrs integration time) and a satellite-borne observatory (assumed integration time: 100 ksec).

For each scattering material, 2 million polarized and 2 million unpolarized photons were simulated. Photons with energies between 10 keV and 80 keV were generated according to the Crab spectrum measured with the Swift Burst Alert Telescope (BAT) telescope [21]:

$$\frac{dN}{dE} = 10.17 \left(\frac{E}{1\,keV}\right)^{-2.15} \text{ph cm}^{-2}\,s^{-1}\,keV^{-1}. \quad (3)$$

For the balloon flight scenario, we account for atmospheric absorption at a residual atmospheric depth of 2.9 gr cm$^{-2}$. For this depth, the transmissivity of the residual atmosphere increases rapidly from 0 to 0.6 in the 20 keV to 60 keV energy range and increases slowly at higher energies (see Fig. 2). We assume a mirror with 40 cm$^2$ effective area from 10 keV to 40 keV. From 40 keV to 80 keV, the effective area drops linearly from 40 cm$^2$ to 0. For the satellite experiment, we assume a 40 times larger effective area than for the balloon experiment.

TABLE I. PHYSICAL CHARACTERISTICS AND DIMENSIONS OF THE DIFFERENT SCATTERING RODS AND CZT DETECTORS FOR EACH CASE

|  | Design 1 | Design 2 | Design 3 | Design 4 |
|---|---|---|---|---|
| Scatterer | Scintillator EJ-200 | Be | LiH | Li |
| Scatterer Diameter | | 1 cm | | |
| Scatterer Length | 14 cm | 9 cm | 18 cm | 32 cm |
| CZT Unit | | 0.5×2×2 cm$^3$ | | |
| CZT Wall Length | 16 cm | 12 cm | 20 cm | 34 cm |

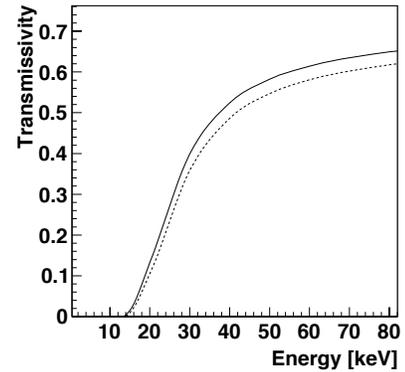

Fig. 2 The transmissivity of the residual atmosphere at a residual atmospheric depth of 2.9 gr cm$^{-2}$.

The incident photons and their secondaries are tracked through the detector volumes while energy deposits and the interaction locations are recorded. For all simulations the instruments were placed in a near-vacuum, similar to a low-Earth orbit environment.

The GEANT4 simulation is followed by a simple detector response simulation. For each event the code computes the energy deposited in the individual detectors and (if applicable) in the pixels of the detectors. A detector signal is used for the analysis if the deposited energy exceeds the energy threshold of the detector. If an event triggers more than one detector element, only the highest energy deposition is used. The azimuthal scattering angle is determined from the triggered CZT pixel location and by assuming the scatter occurred at the center of the low-Z rod.

## V. EXPECTED PERFORMANCE

In this section we discuss the performance achieved with *X-Calibur* using the four different scattering materials. The most important results, including the rate of Compton events for a Crab like source $R_{Crab}$ [Hz], the peak detection efficiency and the energy at which this efficiency is achieved, the modulation factor μ and the minimum detectable polarization MDP, are discussed below and are summarized in Table 2.

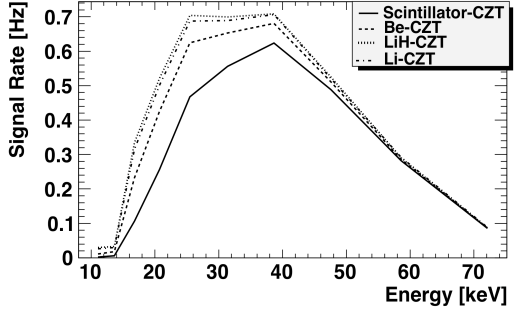

Fig. 3. Comparison of Compton events detection rates. The figure assumes a source with a Crab-like spectrum and flux. The different lines show the results for different scattering materials.

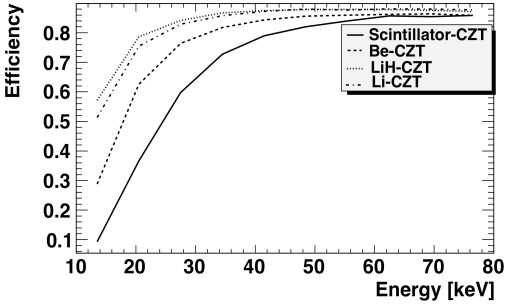

Fig. 4 Comparison of detection efficiencies (number of detected events divided by the number of photons incident on the polarimeter). The different lines show the results for different scattering materials.

### A. Detection Efficiency and Detection Rates

The detection rates of the polarimeter with four different scattering materials for a strong source with a Crab like flux are shown in Fig. 3. The detection efficiencies for the four different scattering materials are shown in Fig. 4. The efficiency is defined here as the fraction of the photons impinging on a detector assembly that trigger the instrument and enter the polarization analysis. The peak efficiencies are listed in Table 2. The simulations show that the lower-Z materials lead to higher rates and efficiencies, especially at the low energies from 10 keV to 50 keV. The scintillator (solid line) gives the lowest rates and efficiencies while LiH (dash-dotted line) gives the highest. The high efficiencies of the X-Calibur design close to 100% can be explained by the fact that all source photons hit the scattering rod close to the optical axis. A large fraction of the photons Compton scatter in the low-Z material, and most of the scattered photons can escape the rod and are detected in the CZT.

### B. Azimuthal Scattering Distributions

Fig. 5. (Top) shows exemplary azimuthal scattering distributions for the Scintillator-CZT configuration for unpolarized and polarized X-ray beams (before correction for non-uniformities). It shows the results for events triggering one or more CZT detectors; The φ-distribution for an unpolarized beam shows some modulation owing to the large pixel size (2.5 mm) and associated binning effects. Before computing the MDP with Eqs. (1) and (2) we correct for binning effects by dividing the polarized distributions by the

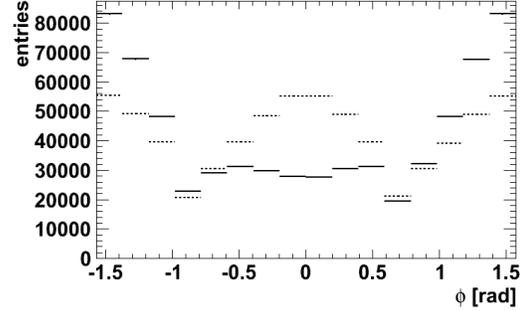

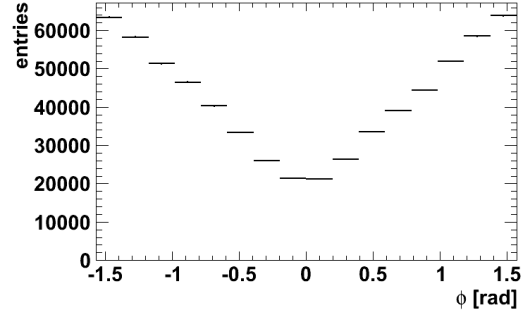

Fig. 5. The upper panel shows distribution of azimuthal scattering angles for a polarized beam (solid lines) and an unpolarized beam (dashed lines). The lower panel shows distributions of azimuthal scattering angles for a polarized X-ray beam after correcting for binning effects.

unpolarized distributions. The correction flattens the φ-distributions of the unpolarized beams and leads to a sinusoidal modulation of the φ-distributions of the polarized beams (see Fig. 5 Bottom). See [1] for a detailed description of the correction procedure and for a study of the validity of Equ. (2) if the correction is used. Typically, modulation factors of ~0.5 are achieved (see Table 2, and Fig. 6, top).

### C. Modulation factor and Minimum Detectable Polarization

Fig. 6, lower panel, shows the sensitivities (MDPs) of the polarimeters with the four different scattering materials for a 6 hrs observation of a source with a Crab like flux and energy spectrum. The MDPs are compiled in Table 2 for the four scattering materials. The LiH scatterer achieves the lowest MDP (1.23%) followed by Li (1.26%), Be (1.46%), and the Scintillator (1.76%). The lower-Z materials perform slightly better, but the scintillator can give a coincidence signal to identify proper Compton events.

We also computed the MDPs of an *X-Calibur* polarimeter for a longer balloon flight and for a satellite mission. Using the scintillator as the Compton scatterer, we derive MDPs of 0.56% for a 60 hrs observation (40 cm$^2$ mirror area), and 0.13% for a 100 ksec observation with a mirror area of 1600 cm$^2$ (see also Table 2).

### VI. SUMMARRY

In this paper we studied the performance of *X-Calibur* on a short 1-day (6 hrs integration time) balloon flight, a longer

TABLE II. COMPARISON OF THE PERFORMANCE OF THE FOUR DIFFERENT SCATTERING MATERIALS

|  | Scintillator-CZT | Be-CZT | LiH-CZT | Li-CZT |
|---|---|---|---|---|
| $R_{Crab}$ (InFOCμS) [Hz] | 3.85 | 4.34 | 4.17 | 4.14 |
| Peak efficiency | 0.86 (77 keV) | 0.86 (70 keV) | 0.87 (77 keV) | 0.88 (77 keV) |
| μ | 0.52 | 0.51 | 0.52 | 0.52 |
| MDP (InFOCμS, 6 hrs) | 1.76% | 1.46% | 1.23% | 1.26% |
| MDP (InFOCμS, 60 hrs) | 0.56% | 0.46% | 0.39% | 0.40% |
| MDP (Satellite Flight, 100 ksec) | 0.13% | 0.11% | 0.09% | 0.09% |

10-day balloon flight (60 hrs integration time) and a 100-ksec observation on a satellite. We considered 4 different Compton scatterers (Scintillator, Be, LiH, Li).

The conclusons from our study can be summarized as follows:
• The lower-Z Compton scattering rods perform slightly better, but the scintillator rod can give a coincidence signal to identify proper Compton events.
• *X-Calibur* combines a detection efficiency close to 100% with a high modulation factor of μ~0.5.
• We derive excellent Minimum Detectable Polarizations for a Crab-like source (using Scintillator as the scattering materials): 1.76% (6 hrs, balloon), 0.56% (60 hrs, balloon) and 0.13% (100 ksec, satellite).

## VII. EXPERIMENTAL DEVELOPMENTS, LABORATORY TESTS AND ONGOING WORK

We are currently working on the optimization of the Scintillator and CZT detectors. Furthermore, we are assembling a full *X-Calibur* polarimeter made of a scintillator rod surrounded by 32 CZT detectors. First measurements are anticipated for January 2011. Furthermore, we are working on detailed background studies and systematic studies.

ACKNOWLEDGMENTS

We acknowledge NASA for support from the APRA program under the grant NNX10AJ56G and support from the high-energy physics division of the DOE. The Washington University group is grateful for discretionary funding of the *X-Calibur* polarimeter by the McDonnell Center for the Space Sciences.

Q.G. thanks the Chinese Scholarship Council from China for the financial support (NO.2009629064) during her stay at Washington University in St Louis.

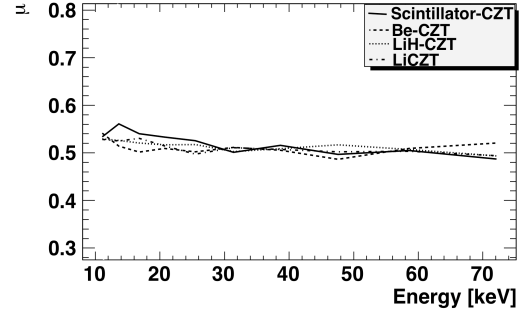

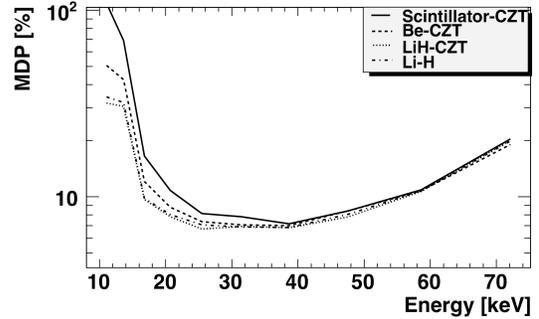

Fig. 6. Comparison of modulation factor (Top) and the MDP (Bottom) for a short (6 hrs) balloon flight. The binning was chosen to have 10 statistically independent bins between 10 keV and 80 keV. The different lines show the results for different scattering materials. The lines between the data points are only shown to guide the eye.